# Evaluating Targeted Mobility Restrictions on COVID-19 Transmission in Seoul: A Metapopulation Modeling Study Using Mobile Phone Data

Yuna Lim[1, †], Jonggul Lee[2, †], and Eunok Jung[1*]

[1] Department of Mathematics, Konkuk University, Seoul, Republic of Korea

[2] Department of Fundamental Research on the Public Agenda, National Institute for Mathematical Sciences

YL, 0009-0005-5475-1461

JL, 0000-0002-5771-3015

EJ, 0000-0002-7411-3134

† Yuna Lim and Jonggul Lee are joint first authors.

This is information for the corresponding author:

Name: **Eunok Jung**

    E-mail: junge@konkuk.ac.kr

    Address: Department of Mathematics, Konkuk University

        120, Neungdong-ro, Gwangjin-gu, Seoul, 143-701, Republic of Korea

    Tel: +82-10-9973-4163

    Fax: +82-2-458-1952

# Abstract

Broad mobility restrictions can help control infectious disease spread, but their socioeconomic costs and the variation in transmission risks by mobility purpose, age group, and spatial connectivity highlight the need for targeted approaches. In this study, we developed an age-structured SEIR metapopulation model for COVID-19 across Seoul's 25 districts, integrating mobile phone-derived origin–destination data. We stratified mobility by age (0–19, 20–59, 60+) and purpose: residential ($H$), school/work ($W$), and other non-routine ($O$). Using 2024 mobility data as a baseline and incorporating pandemic-period (2020–2021) mobility deviations, we investigated counterfactual strategies under various targeting scenarios. Our results showed that $W$ restrictions among adults aged 20–59 produced the highest per-capita reductions in infection. Spatial clustering based on population-adjusted $W$ inflows showed that high-inflow central business districts corresponded to the fast-spreading districts identified in the simulations. Targeting $W$ flows into and within this cluster consistently reduced epidemic size across uncertain seeding locations. Furthermore, weekday-inclusive schedules outperformed weekend-only restrictions. Overall, our findings suggest that although citywide restrictions achieve larger reductions, strategically targeting routine school/work mobility among adults aged 20–59 within fast-spreading clusters can provide substantial epidemiological benefits while reducing broader socioeconomic disruption.

# 1. Introduction

During the early phase of the COVID-19 pandemic, public health systems relied on non-pharmaceutical interventions (NPIs) to curb viral transmission. Multiple studies have shown that reductions in human mobility effectively suppressed epidemic spread [1–3]. However, broad mobility restrictions imposed severe socioeconomic costs; for example, lockdowns led to sharp increases in unemployment, substantial declines in GDP, and prolonged school closures [4–7]. Yet, not all types of mobility contribute equally to transmission because activities, such as commuting to workplaces, attending schools, or engaging in leisure travel, involve distinct contact patterns, durations, and settings, leading to variation in transmission risk [3,8]. If the epidemiological impact of specific mobility types can be quantified, policymakers could design targeted restrictions that prioritize high-risk activities while preserving lower-risk ones, thereby reducing transmission while limiting socio-economic disruption.

The integration of human mobility data into infectious disease modeling accelerated during the COVID-19 pandemic. Pre-pandemic and early-pandemic spatial models relied on commuting flows [9,10] or air travel data to simulate spatial spread [11–13], but these did not systematically distinguish mobility by purpose (e.g., residential, school/work, and other activities). Subsequently, the widespread availability of Google Community Mobility Reports and other related mobility datasets enabled researchers to link temporal variation in movement to epidemic dynamics. During the early phase of the pandemic, reductions in mobility closely tracked declines in transmission [14,15], though this relationship weakened as the pandemic progressed [16,17]. Despite their widespread use, these datasets are limited because they primarily record visit frequencies to specific place categories without capturing the underlying spatial connectivity of trips, including their origins and destinations. Consequently, without origin-destination (OD) information, it is challenging to trace how mobility connects different areas and drives inter-regional transmission.

Korea offers a useful setting to examine how targeted mobility restrictions affect epidemic trajectories because it controlled COVID-19 primarily through testing, contact tracing, and voluntary social distancing, without legally enforced lockdowns [18–20]. Unlike countries that imposed mandatory stay-at-home orders, Korea's response relied on voluntary behavioral changes, which makes it possible to observe how people naturally adjusted their mobility in response to perceived risk. This setting is particularly suited for counterfactual analysis: because lockdowns were never implemented, their potential impact on epidemic trajectories can be estimated through simulation, without the confounding effects of coercive mobility restrictions.

Studies on COVID-19 in Korea have addressed a range of topics, including early transmission forecasting [21,22], optimization of social distancing policies [23,24], and spatiotemporal spread [25]. Notably, Kwon et al. simulated spatial spread across administrative districts using daytime and nighttime population as indirect mobility indicators [25]. However, most existing studies have focused on retrospective evaluations of implemented policies, such as testing, contact tracing, and quarantine alongside social distancing measure. Few have explored what would have happened under policies Korea did not implement, such as lockdowns or purpose-specific travel bans, particularly by using fine-grained OD mobility data.

In this study, we developed an age-structured SEIR metapopulation model for Seoul that stratifies mobility and transmission by age and purpose-specific mobility type, including residential ($H$), school/workplace ($W$), and other non-routine activities ($O$). This model integrates a stratified mobility structure with counterfactual simulations of policies not previously adopted in Korea. By simulating hypothetical mobility restriction scenarios, we quantify the impact of targeted restrictions on epidemic trajectories. Specifically, we compare the epidemic impact of policies varying by age group, mobility purpose, spatial cluster, and timing (weekdays vs. weekends).

This study addresses three key questions. First, how do contributions to transmission differ by age group and mobility purpose, particularly school/work versus non-routine activities? Second, how does Seoul's mobility network structure shape spatial heterogeneity in transmission across districts? Third, how would mobility restriction strategies that vary by age targeting, spatial targeting, and implementation schedule alter epidemic trajectories? Together, these findings aim to provide evidence-based guidance for designing targeted mobility interventions that effectively reduce transmission while minimizing socioeconomic disruption in future pandemics.

## 2. Materials and Methods

### 2.1 Data
The datasets utilized for model calibration comprise the daily number of new confirmed cases across the 25 administrative districts of Seoul and the cumulative number of confirmed cases by age group, obtained from publicly available sources provided by the Seoul Metropolitan Government [26,27]. The study period spans 1 May 2020 to 31 March 2021, covering the second and third waves of COVID-19 in Korea prior to the emergence of the Delta variant. The simulation start date is set to 1 May 2020, marking the period when the proportion of cases in the Seoul Metropolitan Area began to increase significantly, following the initial wave centered in the Daegu–Gyeongsangbuk-do region earlier that year. [28]. The simulation period ends on 31 March 2021 to avoid incorporating the effects of vaccination on transmission dynamics, because the national COVID-19 vaccination program began in late February 2021 in Korea [28,29].

To incorporate human mobility into the model, we use the Seoul mobility dataset, which is publicly available through the Open Data Plaza of the Seoul Metropolitan Government [30]. This dataset is derived from KT mobile phone signals aggregated at base stations and processed to estimate population movements (i.e., number of trips) between origin–destination (OD) pairs. It is provided monthly and includes information on origin/destination, day of the week, arrival time, number of trips, travelers' age and sex, travel duration, and travel type.

We use mobility data from May 2020 to March 2021 and from June, August, and November 2024, restricting the analysis to the 25 districts of Seoul. We treat 2024 mobility as a proxy for pre-pandemic conditions. November 2024 serves as the baseline month for defining mobility proportions and quantifying deviations from the baseline during the epidemic-period, while June and August 2024 are used to represent non-epidemic mobility patterns in counterfactual scenario analyses.

Through preprocessing, we construct 25×25 OD matrices for $P_{ij}^{\omega,a}$, $\sigma_{ij}(t_m)$, and $X_{ij}^{\omega,a}(t_m)$. Their definitions are as follows:

- $P_{ij}^{\omega,a}$: in November 2024, the proportion of movers in age group $a$ on day type $\omega$ (weekday/weekend) who travel from district $i$ to district $j$ among all movers residing in district $i$.

- $\sigma_{ij}(t_m)$: the relative change in the number of individuals moving from district $i$ to district $j$ in month $t_m$ compared with those in the baseline month.

- $X_{ij}^{\omega,a}(t_m)$: among all movers traveling from district $i$ to district $j$ at month $t_m$, stratified by age group $a$ and day type $\omega$ (weekday/weekend), the proportion of movers who are of mobility purpose $X$, where $X \in \{H$: residential, $W$: school or workplace, $O$: other (non-routine)$\}$.

The details of data preprocessing are provided in the electronic supplementary material, section S1.

### 2.2 Mathematical Model
We employ an age-structured deterministic metapopulation model of SARS-CoV-2 transmission across 25 districts in Seoul (Figure 1). In each district $i$ and age group $a$, susceptible individuals $S_i^a$ become exposed $E_i^a$ at rate $\lambda_i^a$, exposed individuals progress to infectious $I_i^a$ after an average latent period $1/\kappa$. Infectious individuals $I_i^a$ move to the removed compartment $R_i^a$ after an average infectious period of $1/\alpha$. The force of infection $\lambda_i^a$ for age group $a$ in district $i$ ($i \in \{1,2,\ldots,25\}$) is defined as:

$$\lambda_i^a = (1-\mu(t_I))\sum_{j=1}^{25}\left(\frac{\sum_{k=1}^{3}\beta_H^{ak}I_j^k}{N_j}H_{ij}^{\omega,a}(t_m) + \frac{\sum_{k=1}^{3}\beta_W^{ak}I_j^k}{N_j}W_{ij}^{\omega,a}(t_m) + \right.$$

$$\left. \frac{\sum_{k=1}^{3} \beta_O^{ak} I_j^k}{N_j} O_{ij}^{\omega,a}(t_m) \right) P_{ij}^{\omega,a} \left(1 + \sigma_{ij}(t_m)\right)$$

Here, $\mu(t_I)$ denotes the time-varying intensity of NPIs (social distancing and personal hygiene) over intervention stages $t_I$, so that $(1 - \mu(t_I))$ captures the corresponding reduction in transmission. The parameter $\beta_X^{ak}$ ($X \in \{H, W, O\}$) represent age- and mobility-purpose-specific transmission rate from infectious individuals in age group $k$. The mobility term $X_{ij}^{\omega,a}(t_m) P_{ij}^{\omega,a}\left(1 + \sigma_{ij}(t_m)\right)$ captures the structure and temporal variation of movements between districts for each mobility purpose: residential ($H$), school/work ($W$), and other non-routine ($O$). By explicitly incorporating observed mobility changes through $X_{ij}^{\omega,a}(t_m), P_{ij}^{\omega,a}$, and $\sigma_{ij}(t_m)$, this formulation attributes transmission variation to a mobility-driven component while capturing the remaining non-mobility-related effect via $\mu(t_I)$, thereby reducing confounding between NPIs and mobility and allowing $\mu(t_I)$ to be estimated more independently of the mobility network.

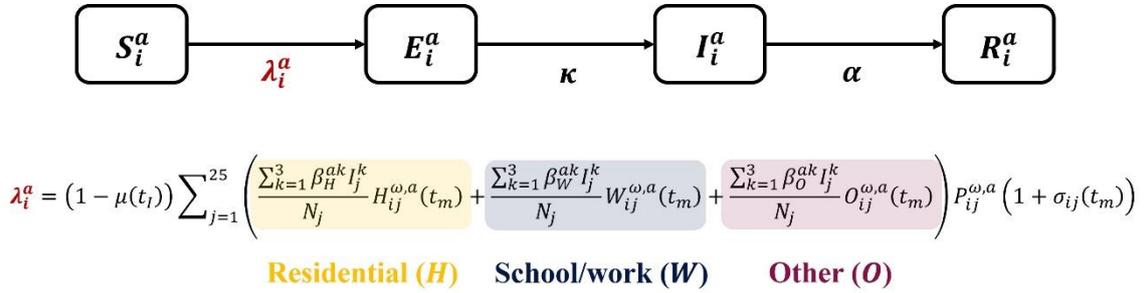

**Figure 1** Age-structured compartment metapopulation model for SARS-CoV-2 transmission in Seoul (25 districts and three age groups). Susceptible individuals ($S_i^a$) are infected at rate $\lambda_i^a$, progress from exposed ($E_i^a$) to infectious ($I_i^a$) after an average latent period $1/\kappa$, and are removed ($R_i^a$) after an average infectious period $1/\alpha$. Each colored term represents a mobility-purpose-specific transmission route: residential ($H$), school/work ($W$), and other non-routine ($O$).

Mobility restrictions are implemented as reduction in $W$ and $O$ mobility, with the restricted portion assumed to be fully substituted into $H$ mobility. In age-targeted scenarios, restrictions are applied across all 25 districts, whereas in cluster-targeted scenarios they are applied to all age groups.

The transmission parameters and the time-varying intervention intensity $\mu(t_I)$ are estimated by simultaneously fitting the model to cumulative cases by district, citywide cumulative cases, and age-specific cumulative cases from May 2020 to March 2021 using a least-squares method (lsqcurvefit). The full differential equations, parameter estimation results, and details of mobility restriction implementation are provided in the electronic supplementary material, section S2, S3, and S5.

### 2.3 Mobility settings for scenario analyses

We consider two mobility settings for the scenario analyses.

First, in a counterfactual non-epidemic mobility setting, we use the OD matrices and mobility-purpose compositions from June and August 2024 to represent typical school-term and vacation-period mobility patterns without epidemic-driven deviations. In this setting, we set $\sigma_{ij}(t_m) = 0$ for all $i$, $j$, and $t_m$. The epidemiological parameters are fixed at values estimated from the COVID-19 period data, and only the mobility network is varied (Result 1).

Second, in an observed epidemic mobility setting, we represent mobility from May 2020 to March

2021 as $P_{ij}^{(\omega,a)}(1 + \sigma_{ij}(t_m))$, thereby incorporating observed increases or decreases in mobility during the epidemic relative to the November 2024 baseline. This setting is used to analyze spatial heterogeneity and inter-cluster restrictions (Result 2) and to evaluate mobility restrictions under different non-pharmaceutical intervention (NPIs) phases (Result 3).

# 3. Results

## 3.1 Age-specific impacts of purpose-specific mobility restrictions

We first use the counterfactual non-epidemic mobility patterns for June and August 2024 to quantify how age-targeted restrictions on school/work ($W$) and other non-routine ($O$) affect the total number of cases in Seoul. Because the mobility patterns differ by month in our data, we select two representative months that capture a major seasonal shift and hold each month's mobility pattern fixed throughout the simulation: June (school term, non-holiday season) and August (school vacation, holiday season). This design allows us to assess whether restriction effects are robust to substantial differences in baseline mobility structure. We focus on per-capita case reductions $A_a$ to compare the effectiveness of restrictions across age groups. The cumulative number of cases in Seoul is defined as

$$CI = \int_0^T \sum_a \sum_i \kappa E_i^a(t) \, dt,$$

where $E_i^a(t)$ is the number of exposed individuals in age group $a$ in district $i$, and $1/\kappa$ is the average latent period. We denote the cumulative cases without any mobility restrictions by $CI^{\text{base}}$, and the cumulative cases when mobility of age group $a$ is restricted by $CI_a^{\text{res}}$. The age-adjusted per-capita case reduction for the restricted age group $a$ is defined as

$$A_a = \frac{CI^{\text{base}} - CI_a^{\text{res}}}{N_a},$$

where $N_a$ is the population size of age group $a$. This quantity represents the reduction in total cases per person in age group $a$ attributable to the restriction.

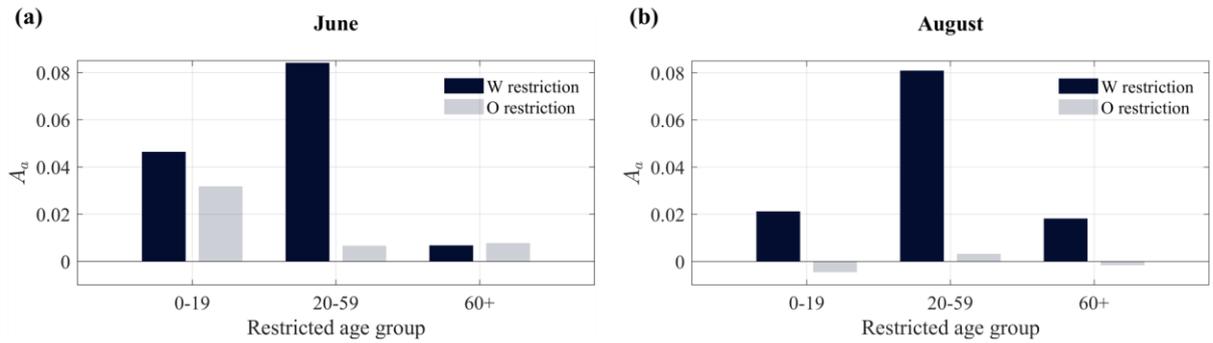

**Figure 2** Age-adjusted per-capita case restriction $A_a$ for each restricted age group under counterfactual non-epidemic mobility: (a) June (school term, non-holiday season) and (b) August (school vacation, holiday season) 2024. Dark blue bars indicate $W$ restrictions; light grey bars indicate $O$ restrictions.

We compare the per-capita case reduction $A_a$ achieved by restricting $W$ or $O$ mobility in each age group (Figure 2). To isolate the effect of mobility interventions, we assume that restrictions start at the onset of the outbreak and that other time-varying elements are held constant (e.g., the estimated NPI intensity $\mu$ and month-specific mobility input are not additionally varied over time). Month-to-month differences between June and August reflect changes in the mobility network and the resulting $W/H/O$ composition of the force of infection (electronic supplementary material, Figure S2).

Across both June and August mobility settings, $W$ restrictions consistently yield larger per-capita reductions than $O$ restrictions. In particular, restricting $W$ mobility among adults aged 20–59 produces the largest $A_a$ in both months, suggesting that routine school/work mobility in this age group contributes substantially to overall transmission. $W$ restrictions targeting ages 0–19 and 60+ also reduce cases, but their per-capita effects are smaller than those for ages 20–59.

By contrast, $O$ restrictions produce $A_a$ values close to zero in most scenarios, and in some August scenarios they become slightly negative for ages 0–19 and 60+, indicating a small increase in cumulative cases relative to the baseline. This limited impact of $O$ restrictions is consistent with the mobility-purpose decomposition of the force of infection (electronic supplementary material, Figure S3), suggesting that reductions in $O$-related movement can be offset by contributions from other mobility components. Overall, these results are robust across the two representative months: interventions targeting routine school/work mobility, especially among adults aged 20–59, consistently achieve the largest reductions in cumulative cases.

Figure 3 illustrates the dependence of per-capita case reduction $A_{20-59}$ on the timing of mobility restrictions in adults aged 20–59 years. For $W$ restrictions in this age group, delaying the start date after the initial outbreak weakens the reduction effect under both the June and August mobility patterns. In contrast, $O$ restrictions in adults aged 20–59 years show no consistent dependence on the start date, and their effects remain small in all timing scenarios.

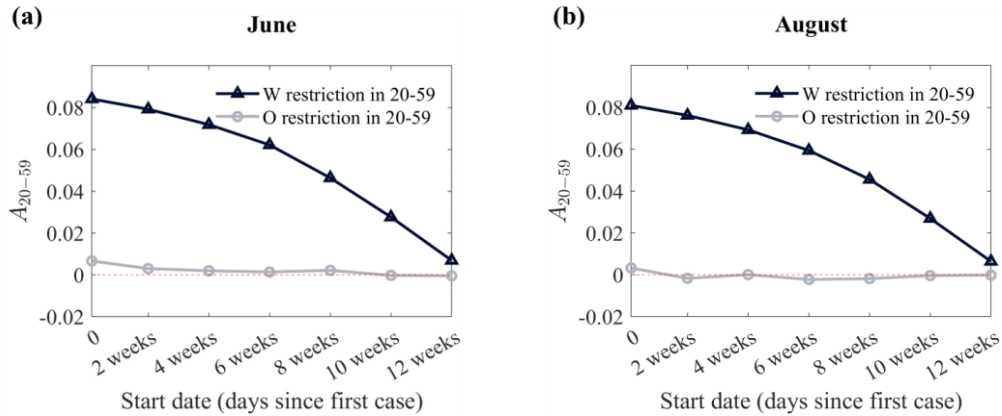

**Figure 3** Dependence of the age-adjusted per-capita case reduction $A_{20-59}$ on the start date of mobility restrictions targeting adults aged 20–59 years under counterfactual non-epidemic mobility: (a) June and (b) August 2024. Filled triangles indicate $W$ restrictions and open circles indicate $O$ restrictions. The x-axis denotes the restriction start date (weeks since the first case).

Taken together, these findings indicates that $W$ restrictions in adults aged 20–59 are most effective when early, but remain effective even with moderate delays in implementation.

**3.2 Spatial heterogeneity and the robust effects of cluster-targeted $W$ mobility restrictions**

Using the observed epidemic-period mobility, $P_{ij}^{\omega,a}(1 + \sigma_{ij}(t_m))$, we examine whether Seoul exhibits spatial heterogeneity in epidemic growth and whether such heterogeneity aligns with the pattern of routine school/work ($W$) mobility. We then assess how mobility-based spatial targeting performs when implementing $W$ mobility restrictions under uncertainty in the outbreak source.

To characterize spatial heterogeneity, we simulate the total cumulative number of cases in Seoul under hypothetical scenarios in which the initial outbreak is seeded in each of the 25 districts, while keeping all other parameters and initial conditions fixed (Figure 4a). The y-axis represents the source district, the x-axis represents time, and the color scale indicates the total cumulative cases in Seoul resulting from an outbreak seeded in that district. The cyan box highlights the trajectory of cumulative cases observed in the actual case data.

Based on whether the simulated cumulative case trajectories reach the cyan-boxed range at early or later times, we categorize districts into two groups with relatively faster or slower epidemic growth. Here mobility is modeled as $P_{ij}^{(\omega,a)}(1 + \sigma_{ij}(t_m))$, which incorporates observed mobility changes during 2020–2021 relative to the November 2024 baseline.

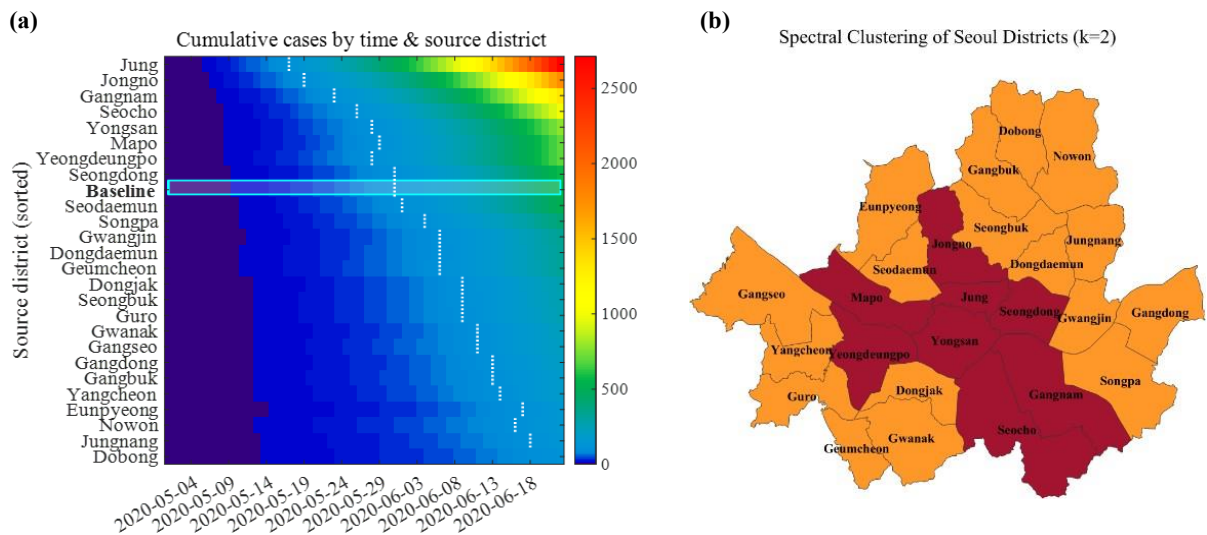

**Figure 4** (a) Total cumulative number of cases in Seoul under hypothetical scenarios where the initial outbreak is seeded in each district; the cyan box indicates the cumulative cases observed in the data. The dotted vertical line in each district indicates the day when the cumulative number of cases reaches 100. (b) Spatial distribution of the two mobility clusters on the map of Seoul.

We next examine whether this fast-slow dichotomy corresponds to routine mobility patterns. For each month from May 2020 to March 2021, we construct population-adjusted vectors of $W$ inflows into each district and apply spectral clustering to the similarity of these $W$ inflow time series. Details of the spectral clustering procedure are provided in the electronic supplementary material, section S8. This analysis partitions Seoul into two mobility clusters (Figure 4b):

• Cluster 1: central business and commercial districts (Junggu, Jongnogu, Gangnamgu, Seochogu, Yongsangu, Mapogu, Yeongdeungpogu, Seongdonggu), shown in red in panel (b)
• Cluster 2: surrounding residential districts, shown in yellow in panel (b)

Notably, the mobility-based clustering matches the fast-slow grouping obtained from the seeding simulation (Figure 4a). Taken together, these results indicate that districts with higher routine $W$ inflows form a coherent cluster and that this cluster is closely aligned with districts exhibiting faster epidemic growth in the seeding analysis.

We then evaluate how this cluster structure can be used for spatially targeted restrictions when the outbreak source is uncertain. Specifically, we seed the initial outbreak in each of Seoul's 25 districts and compare two $W$ restriction strategies applied to all ages: restricting $W$ mobility into and within Cluster 1 versus Cluster 2 (Figure 5). Each point corresponds to one assumed source district, and the boxplots summarize the distribution of total cumulative cases across source districts. The dashed–dotted line denotes the median cumulative cases under the no mobility restriction baseline.

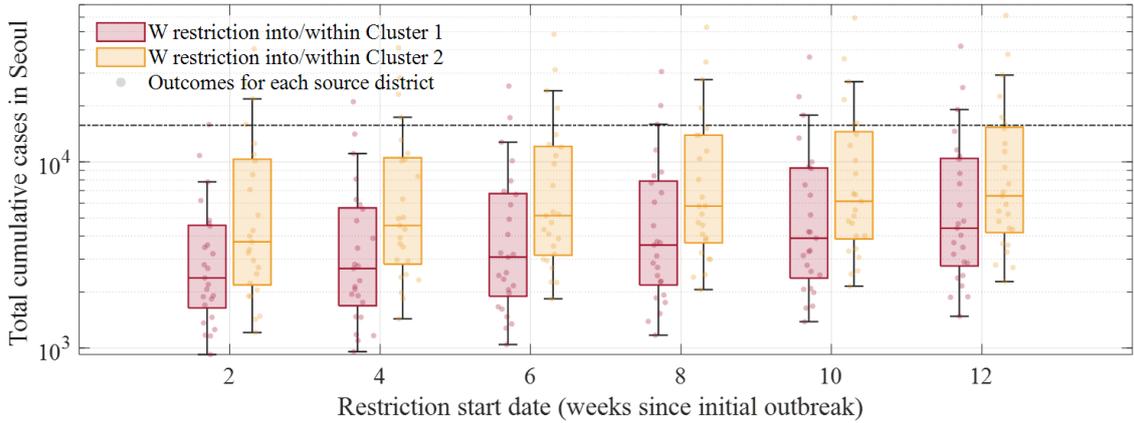

**Figure 5** Boxplots of total cumulative cases in Seoul under $W$ mobility restrictions targeting Cluster 1 versus Cluster 2, computed across outbreaks seeded in each of the 25 source districts. The x-axis indicates the restriction start date (weeks since the initial outbreak) and the y-axis shows total cumulative cases (log scale). Points show outcomes for each source district (red: $W$ restriction into/within Cluster 1; yellow: $W$ restriction into/within Cluster 2). The dashed-dotted line denotes the median total cases under no mobility restriction across the 25 source-district scenarios.

Across seeding locations, epidemic size varies substantially, highlighting the importance of spatial heterogeneity and source uncertainty. Delaying the start of restrictions increases cumulative cases under both targeting strategies. Nevertheless, across all source districts and restriction start dates, the median and interquartile range of total cumulative cases are consistently lower when targeting Cluster 1 than when targeting Cluster 2. These findings indicate that restricting routine $W$ flows into and within the central business/commercial cluster provides a robust reduction in epidemic size regardless of the outbreak's source district.

### 3.3 Scenario-based comparison of $W$ mobility restrictions under different NPI phases

Finally, combining the estimated time-varying NPI intensity $\mu(t_I)$ with observed epidemic-period mobility $P_{ij}^{(\omega,a)}(1 + \sigma_{ij}(t_m))$, we evaluate several $W$ restriction strategies implemented alongside social distancing measures in Korea under different NPI phases and application schedules. Based on Results 1 and 2, we define four $W$ restriction scenarios by crossing age targeting (all ages vs 20–59 years) and spatial targeting (citywide vs Cluster 1):

(1) all age groups in all districts of Seoul

(2) adults aged 20–59 years in all districts of Seoul

(3) all age groups for $W$ flows into and within Cluster 1

(4) adults aged 20–59 years for $W$ flows into and within Cluster 1

Figure 6a shows the estimated NPI intensity, $\mu(t_I)$, which reflects the strength of social distancing and personal hygiene. The estimated $\mu(t_I)$ follows the temporal pattern of the social distancing levels enforced in Korea during the COVID-19 epidemic. A discrepancy is observed between 8 August and 30 August 2020, when the estimated $\mu(t_I)$ is lower than expected despite stricter interventions compared with the preceding period (24 June – 8 August 2020). This deviation is likely attributable to mass infection events, including the Liberation Day rally on August 15 and outbreaks linked to a religious group [31,32].

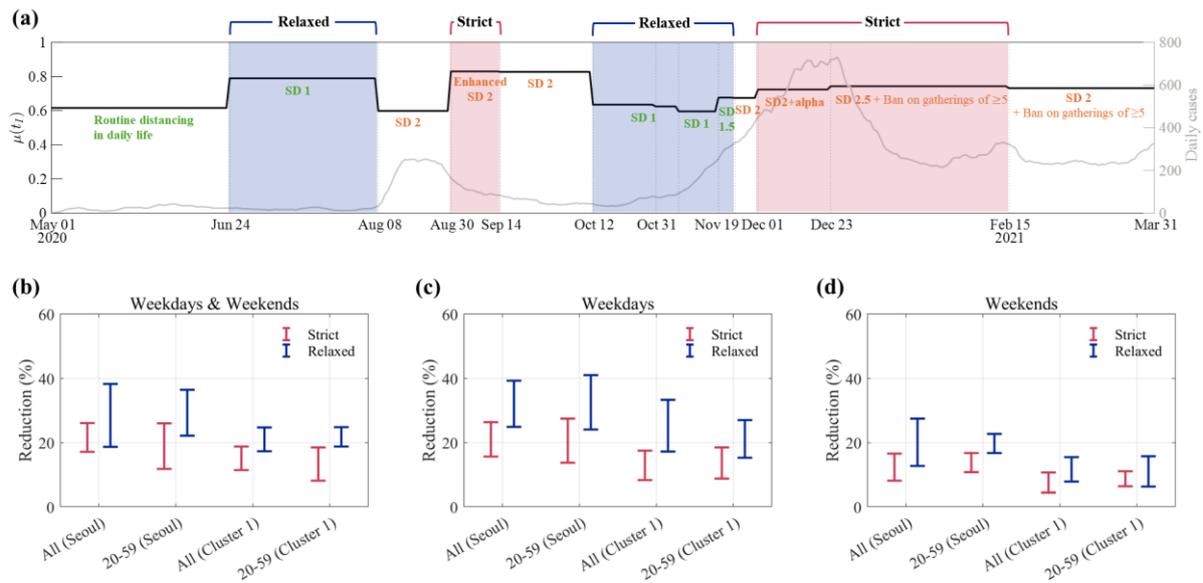

**Figure 6** Effects of $W$ mobility restrictions under different NPI phases and application schedules. (a) Estimated NPI intensity, $\mu(t_I)$ (tick black line), with Relaxed (blue shading area) and Strict (red shading area) periods, and the 7-day moving average of daily cases (grey line). (b–d) Percentage reduction in total cumulative cases when $W$ restrictions are implemented during relaxed (blue) or strict (red) periods under three schedules: (b) weekdays & weekends, (c) weekdays only, and (d) weekends only. Vertical lines indicate the minimum-maximum range across $W$ restriction levels from 50% to 100%.

Using $\mu(t_I)$ and actual intervention levels, we classify the simulation period (excluding the routine daily life distancing phase: 1 May – 24 June 2020) into two categories:

- *Relaxed period* (blue-shaded area): Relatively less stringent social distancing (24 June – 8 August 2020; 12 October – 24 November 2020; total 88 days)

- *Strict period* (red-shaded area): Relatively more stringent social distancing (30 August – 14 September 2020; 1 December 2020 – 15 February 2021; total 90 days)

Figure 6b–d show the percentage reduction in cumulative cases when $W$ mobility restrictions are imposed during each period under three schedules (b: Weekdays & Weekends, c: Weekdays only, d: Weekends only). For comparability, restrictions are applied for a fixed total of 26 days in both periods, matching the number of weekend days in each period. Across panels b–d, $W$ mobility restrictions yield larger percentage decreases in cumulative cases when implemented during the Relaxed period than during the Strict period. Moreover, over the same 26-day duration, weekday-only restrictions produce reductions in cumulative cases comparable to those under the weekdays & weekends schedule in both periods and across targeting scenarios, whereas weekend-only restrictions consistently result in smaller percentage decreases in cumulative cases.

Across panels b–d, within the same geographic strategy, age targeting has a limited effect, with reductions showing limited variation across age groups. In contrast, geographic targeting is significantly more impact. Seoul-wide $W$ restrictions consistently achieve larger reductions than Cluster 1 targeting under the same age-targeting scenario. Still, targeting Cluster 1 for working-age adults (20–59) produces meaningful reductions across schedules; for example, under the weekday-only schedule (Figure 6c), it reduces cumulative cases by 9–19% during the Strict phase and 15–27% during the Relaxed phase. Overall, the impact of $W$ restrictions is shaped more by where they are applied than by which age group is targeted, and it also varies by NPI phase (Strict vs Relaxed) and application schedule (weekday-inclusive vs weekend-only).

# 4. Discussion

This study uses a district-level metapopulation framework in which mobility is stratified by purpose, age group, and district to investigate the differential contributions to transmission and the effectiveness of targeted restrictions in a highly connected metropolitan setting. By separating mobility into school/workplace ($W$), residential ($H$), and other non-routine ($O$) components, we addressed three questions. First, regarding the differential contributions of mobility purpose and age group, $W$ restrictions targeting adults aged 20–59 consistently produced the largest per-capita reductions, whereas $O$ restrictions yielded small and inconsistent effects, indicating that intervention effectiveness depends on whether restrictions disrupt key routine pathways rather than on the overall magnitude of mobility reduction. Second, regarding spatial heterogeneity, districts with high routine $W$ inflows formed a distinct cluster aligned with faster-spreading districts, suggesting that the spatial structure of routine mobility shapes epidemic growth across the city. Third, regarding restriction strategies, the epidemiological impact of $W$ restrictions varied with concurrent social distancing intensity and implementation schedule, with weekday-focused restrictions achieving benefits comparable to full-week schedules under a fixed number of total restricted days. Together, these findings suggest that targeting routine $W$ mobility, prioritizing high-connectivity districts, and aligning restrictions with periods of lower social distancing intensity can achieve meaningful epidemiological gains while limiting the need for broad, population-wide measures.

Our findings highlight the importance of distinguishing mobility components when designing and evaluating restrictions, rather than treating mobility as a single undifferentiated driver. The differential effects arise because each component reflects distinct contact structures and inter-district connectivity. Routine school/workplace ($W$) mobility involves daily, repeated trips along fixed origin–destination routes, creating persistent inter-district mixing that can sustain transmission across the city. By contrast, non-routine ($O$) mobility tends to be more dispersed and irregular, so its reduction produced small and inconsistent effects in our simulations. This mechanistic distinction is supported by evidence that job commuting is a primary spatial transmission channel. Specifically, a doubling of cases in commuting-connected regions has been associated with up to a 20% increase in local cases, and this effect diminished once mobility restrictions were in place [33]. Similarly, inter-district mobility, particularly among working-age adults, has been shown to be more strongly associated with COVID-19 transmission than intra-district movement in a comparable metropolitan setting [34], consistent with our finding that $W$ restrictions targeting adults aged 20–59 produced the largest per-capita reductions. These patterns align with broader evidence that Rt is more strongly linked to activity in workplaces, transit, and retail settings than in parks, underscoring the policy value of targeting routine contact settings. The spatial implications of this routine mobility structure are examined next [35].

If routine $W$ mobility is a key transmission pathway, the next question is where such flows are concentrated and whether their spatial structure shapes epidemic growth across districts. Our simulations showed that central business and commercial districts with high routine $W$ inflows were closely aligned with faster-spreading districts, suggesting that these districts play a disproportionate role in shaping epidemic growth across the city. This pattern is consistent with recent evidence of urban superspreading, in which a small fraction of neighborhoods accounts for a large share of subsequent infections, attributed to the joint effect of uneven population distribution and amplified mobility flows [36]. Targeting $W$ flows into and within the high-inflow cluster yielded larger reductions in epidemic size than targeting other districts, and this advantage held across different assumed outbreak sources. Although citywide $W$ restrictions produced larger reductions overall, targeting the high-inflow cluster for working-age adults still yielded meaningful reductions, suggesting that spatially focused strategies may remain effective when broader restrictions are difficult to implement.

Our results also suggest that the impact of mobility restrictions is context dependent, varying with the intensity of concurrent social distancing measures and implementation schedule. When social distancing was already stringent and baseline transmission was suppressed, mobility restrictions tended to provide smaller incremental gains; conversely, during more relaxed social distancing phases, they contributed to larger marginal reductions. This pattern is consistent with cross-country evidence that the mobility–transmission association weakens where facial covering rules are implemented, suggesting that when protective behaviors are widespread, the marginal contribution of further mobility reductions may be limited [37]. Beyond timing, the schedule of restrictions also shaped their impact: under the same total number of restricted days, weekday-focused schedules achieved reductions comparable to full-week schedules, whereas weekend-only schedules consistently produced smaller reductions. Together, these results suggest that mobility restrictions are most effective when implemented during periods of lower social distancing intensity and structured around routine weekday contact patterns.

Across these three findings, our results suggest a what–where–when–how framework for designing targeted mobility interventions in metropolitan areas. Restrictions are most effective when they target routine school/workplace mobility (what), are concentrated on high-inflow, high-connectivity districts (where), are timed to periods when social distancing intensity is relatively low (when), and are scheduled around weekday contact patterns when the number of restricted days is limited (how). Applying these four dimensions jointly may help policymakers achieve meaningful epidemiological gains without resorting to broad, population-wide restrictions

Several limitations should be noted when interpreting these results. First, the mobility network covers only trips within Seoul's 25 districts and does not capture inter-regional or international movements that may seed infections or shape spatial spread; the hub-based targeting effects identified here may therefore differ in magnitude when applied to broader metropolitan or multi-city networks. Second, mobility restrictions are modeled as idealized, uniform reductions in $W$ and $O$ flows, with the restricted portion fully substituted into residential $H$ mobility. Because imperfect adherence and heterogeneous compliance are not explicitly modeled, our estimates represent upper bounds on achievable reductions under real-world conditions. Third, we treat intervention schedules as externally imposed and do not model behavioral feedback or fatigue. In practice, prolonged or frequently repeated restrictions may alter individual behavior in ways our framework does not capture. Accordingly, our estimates should be interpreted as illustrating the relative epidemiological impact of alternative targeting strategies under the assumed conditions, rather than as precise predictions of real-world outcomes.

Nonetheless, by decomposing transmission by mobility purpose, age, and district within a network model, the analytical framework developed here provides a reusable structure for evaluating targeted intervention strategies in other metropolitan settings. While the specific effects identified are calibrated to Seoul's mobility network and COVID-19 dynamics, the broader principle that intervention effectiveness is shaped by routine contact pathways, spatial connectivity, and policy timing is likely to generalize across dense urban environments. Extending this framework to multi-city or nationwide mobility networks remains an important direction for future work, as inter-regional movements may modify both the magnitude and spatial distribution of targeted restriction effects. In conclusion, this study demonstrates that decomposing mobility by purpose, age, and spatial structure provides a principled basis for designing targeted interventions that achieve meaningful reductions in epidemic spread while minimizing the socioeconomic burden of broad, population-wide restrictions.

## Ethics

## Data accessibility

## Declaration of AI use

## Authors' contributions

Y.L.: Conceptualization, Data curation, Formal analysis, Investigation, Methodology, Software, Validation, Visualization, Writing – original draft, Writing – review & editing; J.L.: Conceptualization, Data curation, Formal analysis, Investigation, Methodology, Supervision, Writing – original draft, Writing – review & editing; E.J.: Conceptualization, Funding acquisition, Project administration, Resources, Supervision, Writing – review & editing.

## Conflict of interest declaration

## Funding

This research was supported by 'The Government-wide R&D to Advance Infectious Disease Prevention and Control', Republic of Korea (grant number: HG23C1629). This paper was also supported by the Korea National Research Foundation (NRF) grant funded by the Korean government (MEST) (NRF-2021R1A2C100448711). This work was supported by the National Institute for Mathematical Sciences (NIMS) grant funded by the Korean government (No. NIMS-B26730000).